\documentclass[sigconf]{acmart}
\AtBeginDocument{%
  \providecommand\BibTeX{{%
    \normalfont B\kern-0.5em{\scshape i\kern-0.25em b}\kern-0.8em\TeX}}}

\setcopyright{acmcopyright}
\copyrightyear{2021}
\acmYear{2021}
\acmDOI{}

\acmConference[JCDL '21]{The ACM/IEEE Joint Conference on Digital Libraries (JCDL)}{Sept. 27--30, 2021}{Online}
\acmBooktitle{The ACM/IEEE Joint Conference on Digital Libraries (JCDL), Sept. 27--30, 2021, Online}

\usepackage{hologo}
\usepackage{booktabs}
\usepackage{multirow}

\begin{document}

\title{ACM-CR: A Manually Annotated Test Collection for Citation Recommendation}

\author{Florian Boudin}
\email{florian.boudin@univ-nantes.fr}
\orcid{0000-0001-5849-2261}
\affiliation{%
  \institution{LS2N, Universit\'{e} de Nantes}
  \city{Nantes}
  \country{France}
}

\begin{abstract}
Citation recommendation is intended to assist researchers in the process of searching for relevant papers to cite by recommending appropriate citations for a given input text.
Existing test collections for this task are noisy and unreliable since they are built automatically from parsed PDF papers. 
In this paper, we present our ongoing effort at creating a publicly available, manually annotated test collection for citation recommendation.
We also conduct a series of experiments to evaluate the effectiveness of content-based baseline models on the test collection, providing results for future work to improve upon.
Our test collection and code to replicate experiments are available at \url{https://github.com/boudinfl/acm-cr}
\end{abstract}




\maketitle

\section{Introduction}

Citing has always been an integral part of academic research, whether it is for backing up claims or for referring to previous work of relevance.
Yet, citing properly is becoming increasingly difficult and time-consuming as the volume of published research continues to grow exponentially~\cite{nature-2009-credit}.
Researchers are under constant pressure to publish their findings, but have less and less time to browse the literature to retrieve relevant papers to cite.
This has motivated an active line of research on citation recommendation systems (see~\cite{Farber2020} for a recent survey), whose goal is to relieve the researchers from performing this task by recommending appropriate citations for a given text.

\begin{figure}[h!]
    \centering
    \resizebox{.475\textwidth}{!}{\includegraphics{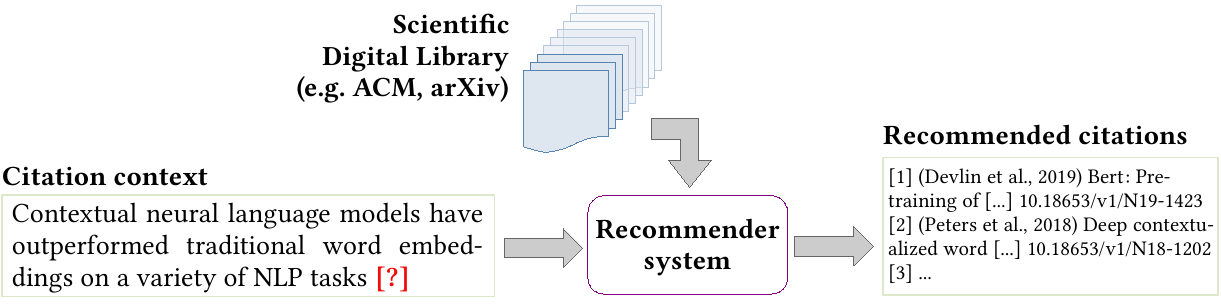}}
    \vspace{-1.5em}%
    \caption{Illustration of the citation recommendation task.}
    \label{fig:citation-recommendation}
    \vspace{-.5em}%
\end{figure}

More precisely, citation recommendation is the task of finding relevant citations for a given \emph{citation context}, that is, a text passage (e.g.~a sentence or a paragraph) within a document (See Figure~\ref{fig:citation-recommendation}).\footnote{This is not to be confused with paper recommendation, which is the task of recommending documents to the user that are worth reading~\cite{beel-2016-paper}.}
Citation recommendation, also referred to as \emph{context-aware citation recommendation} or \emph{local citation recommendation} in previous work, is often viewed as a retrieval task where, given a citation context (query), the task is to retrieve the most relevant citations (documents) from a collection of scientific texts.
The benefits of doing so are two-fold: 1) well-known retrieval models can be readily applied to the task at hand, and 2) the effectiveness of citation recommendation systems can be evaluated offline through \emph{test collections}.
%

\vspace{-.5em}%

Existing test collections for citation recommendation are built automatically by extracting citation contexts (queries) and cited reference(s) (relevant judgments) from a collection of scientific papers.
As a result, they are known to be quite noisy and unreliable due to errors in citation parsing and/or PDF to text conversion~\cite{Farber2020}.
Other reported problems include muddled citation contexts (roughly defined as fixed-size windows of characters around citations), and incomplete metadata information of papers.
In this work, we focus on addressing these problems and present our ongoing effort to create ACM-CR, a new publicly available, manually annotated test collection for citation recommendation.

\section{The ACM-CR test collection}

As a first step, we assembled a sizeable collection of documents by collecting bibliographic records of scientific papers (\hologo{BibTeX} entries) from the ACM Digital Library.
Our document collection currently holds 114,882 records of scientific papers on topics related to Information Retrieval (IR) and adjacent research fields (e.g.~machine learning, databases or data mining).\footnote{We use the SIGs IR, KDD, CHI, WEB and MOD sponsored conferences and related journals as a means to filter articles.}
Each record carries a Digital Object Identifier (DOI), and most of them (91\%) provide paper abstracts which are the primary units of indexing in scientific literature search engines~\cite{10.1145/3295750.3298953}.

To create queries and relevance judgments, we collected 50 open-access scientific papers published at top-tier IR conferences in 2020\footnote{Venues are SIGIR, CHIIR, ICTIR and WSDM.}, from which we manually extracted citation contexts and cited references.
For citation contexts, we extracted full paragraphs that include citations from the introduction and related work sections of each paper (other sections being less suitable for the task~\cite{jurgens-etal-2018-measuring}).
We then took a step further and removed end-of-line hyphens, split paragraphs into sentences and anonymized citations that reveal their author's identity (i.e.~``Smith et al. [1] proposed'' is reformulated as ``[1] proposed'').
For cited references, we mapped each reference to its DOI by querying Google Scholar\footnote{\url{https://scholar.google.com/}}.
We paid a particular attention to preprint references in case they have been superseded by refereed publications.
For those without DOIs but available online, we indicated the URLs to get their PDF files for later use.
The papers we collected have 31.8 cited references on average, half of which occur in our document collection.
Annotating a single paper took about an hour on average, all annotations being encoded in XML format.

Table~\ref{tab:stats-citation-contexts} presents some statistics about the test collection.
Overall, we extracted 341 citation contexts (paragraphs) from which 269 can be used as queries for evaluation because they have cited references that appear in our document collection (underlined in Table~\ref{tab:stats-citation-contexts}).
%
It should be noted that our citation contexts are on average twice as long as the 400 characters used in previous work, and thus we might expect better retrieval results.
On a lower level of granularity, we annotated 837 out of the 1,800 sentences that make up the 341 citation contexts, each of which has 2 cited references on average.
Again, a smaller portion of these annotated sentences (552) have cited references that appear in our document collection and can be used as queries.
Here, the idea is to provide two different types of queries (paragraphs and sentences) to investigate the effect of citation context length on retrieval performance.

\begin{table}[h!]
    \centering
    \begin{tabular}{l | rr | rrr}
    \toprule
        \textbf{Context} &
        \textbf{\#nb} &
        \textbf{\#queries} &
        \textbf{\#tokens} &
        \textbf{\#char.} &
        \textbf{\#cit.} 
        \\
    \midrule
        Paragraphs & 341 & 269 & 145.4 & 809 & 5.1  \\
        $\llcorner$ Sentences  & 837 & 552 & 30.9 & 164 & 2.1   \\
    \bottomrule
    \end{tabular}
    \vspace{.5em}%
    \caption{Statistics of the test collection. The average number of tokens, characters and citations per context are reported. }
    \label{tab:stats-citation-contexts}
    \vspace{-2em}%
\end{table}

\section{Experiments}

We report a series of experiments to evaluate the effectiveness of content-based citation recommendation models on the introduced test collection.
It serves two main purposes: 1) to perform an initial sanity check on the test collection, and 2) to provide baseline results for future work to improve upon.
The first model we consider is BM25, a standard \textit{ad-hoc} retrieval model that is commonly used as baseline in citation recommendation.
Specifically, we use the implementation of BM25 from the Anserini open-source IR toolkit~\cite{Yang:2017:AEU:3077136.3080721} with the default parameters.
For the second model, we adopt a two-stage neural document ranking approach inspired by~\cite{nogueira2020evaluating}, and use SciBERT~\cite{scibert} to re-rank the top-20 documents retrieved by BM25.
%
We apply the uncased model\footnote{\url{https://github.com/allenai/scibert}} without fine tuning and re-rank documents against citation contexts by computing the cosine similarity between their hidden representations.
We evaluate effectiveness of the models in terms of recall and nDCG at the top 10 recommendations as recommended in~\cite{Farber2020}.
We use the Student’s paired t-test to assess statistical significance of our retrieval results at $p<0.05$.

\begin{table}[ht!]
    \centering
    \begin{tabular}{l|rr|rr}
        \toprule
        \multirow{2}{*}{\textbf{Model}} & 
        \multicolumn{2}{c|}{\textbf{Paragraphs}} &
        \multicolumn{2}{c}{\textbf{Sentences}}
        \\
        ~ & \small{R@10} & \small{nDCG@10} & \small{R@10} & \small{nDCG@10} \\
        \midrule
        BM25 &  33.58 & 28.27 & 39.03 & 31.56 \\
        BM25+SciBERT & 34.40 & 29.37$^\dagger$ & 40.30$^\dagger$ & 32.31$^\dagger$\\
        \bottomrule
    \end{tabular}
    \vspace{.5em}%
    \caption{Retrieval effectiveness of content-based citation recommendation models. $\dagger$ indicates significance over BM25.}
    \label{tab:results}
    \vspace{-2em}%
\end{table}



Results are presented in Table~\ref{tab:results}.
We observe noticeably higher scores for the BM25 model in contrast to prior work (e.g.~$\approx+10\%$ in comparison to~\cite{10.1145/3383583.3398534}), which we attribute to the high quality of the manually extracted citation contexts.
Re-ranking using SciBERT increases the retrieval effectiveness, with statistically significant improvements in three out of four cases.
The impact of re-ranking is more important on sentences, which seems reasonable since short queries are prone to vocabulary mismatch issues.
%

\section{Conclusion and Future Work}

In this paper, we described our progress in creating ACM-CR, the first manually annotated test collection for citation recommendation.
We hope that this resource will provide a useful benchmark for future studies, and help us gain better insights on the effectiveness of citation recommendation models. 
For future work, we plan to collect and annotate more papers, while also exploring two directions for improving our test collection.
The first one will be to obtain the full-text of all the open-access articles in our document collection, and to construct the underlying citation graph on which many citation recommendation models rely on.
The second direction concerns the incomplete nature of the extracted relevance judgments, i.e.~that do not include uncited, yet relevant papers.
Here, we will investigate how co-citation instances can be used to overcome the sparseness of the relevance judgments.






\begin{acks}
This work was supported by the French National Research Agency (ANR) through the DELICES project (ANR-19-CE38-0005-01).
\end{acks}

\bibliography{anthology,custom}
\bibliographystyle{ACM-Reference-Format}

\end{document}